\documentstyle[11pt,aasms4]{article}

\begin{document}

\title{An Interferometric Search for Bright Companions to 51~Pegasi}
\author{A.F.~Boden\altaffilmark{1},
	G.T.~van Belle\altaffilmark{1},
	M.M.~Colavita\altaffilmark{1},
	P.J. Dumont\altaffilmark{1},
	J.~Gubler\altaffilmark{2},
	C.D.~Koresko\altaffilmark{3},
	S.R.~Kulkarni\altaffilmark{3},
	B.F.~Lane\altaffilmark{3},
	D.W.~Mobley\altaffilmark{1},
	M.~Shao\altaffilmark{1},
	J.K.~Wallace\altaffilmark{1}
	(The PTI Collaboration)}
\altaffiltext{1}{Jet Propulsion Laboratory, California Institute of Technology}
\altaffiltext{2}{University of California, San Diego}
\altaffiltext{3}{Palomar Observatory, California Institute of Technology}
\authoremail{bode@huey.jpl.nasa.gov}

\begin{abstract}
We report on a near-infrared, long-baseline interferometric search for
luminous companions to the star 51~Pegasi conducted with the Palomar
Testbed Interferometer.  Our data is completely consistent with a
single-star hypothesis.  We find no evidence to suggest a luminous
companion to 51~Pegasi, and can exclude a companion brighter than a
$\Delta$K of 4.27 at the 99\% confidence level for the 4.2-day orbital
period indicated by spectroscopic measurements.  This $\Delta$K
corresponds to an upper limit in the companion M$_K$ of 7.30, in turn
implying a main-sequence companion mass less than 0.22 M$_{\sun}$.
\end{abstract}


\section{Introduction}

The recent inference of a planetary-mass gravitational companion to
the star 51~Pegasi (HD 217014) from apparent radial velocity variation
by Mayor \& Queloz (1995) has subjected this otherwise unremarkable
star to remarkable scrutiny.  The Mayor and Queloz result was quickly
verified by several groups with similar or higher-resolution
spectroscopic techniques (c.f.~\cite{Marcy97}).  However, there has
been no other evidence for a companion, e.g.~precision photometric
monitoring has failed to show evidence for eclipses (\cite{Henry97}),
and there is a significant lack of x-ray flux from the system compared
to binary systems with similar periods (\cite{Pravdo96}).
Further, 51~Peg's G5V spectral classification has become mildly controversial
(e.g.~\cite{Houk95}, who argues for a G2-3V), as has its physical size
(e.g.~\cite{Hatzes97,Henry97}).

A planetary-mass companion in a 4.2 day orbit around a solar-mass
51~Peg would have an orbital semi-major axis of approximately 0.05 AU
(\cite{Marcy97}), slightly more if the companion were more massive.
At a distance of 15.4 $\pm$ 0.2 pc (\cite{Perryman96}), the
approximate maximum primary-companion angular separation would be 3.5
millarcseconds (mas).  Such an angular separation is well below
resolution limits for current conventional imaging technology, but is
accessible to optical and near-infrared interferometry.  As only the
lower mass limit is set by the spectroscopic results, it is possible
the companion is significantly more massive -- perhaps even a low-mass
star.  We have therefore studied 51~Peg with the Palomar Testbed
Interferometer (PTI) in an attempt to detect the putative companion if
it is indeed sufficiently luminous.  PTI is a 110m-baseline
interferometer operating at K-band (2 -- 2.4 $\mu$m) located at
Palomar Observatory, and described in detail elsewhere
(\cite{Colavita94}).  The minimum PTI fringe spacing is roughly 4 mas
at the sky position of 51~Peg, making a (sufficiently) luminous
companion readily detectable.

\section{Experiment Design}

The observable used for these observations is the fringe contrast or
{\em visibility} (squared) of an observed brightness distribution on
the sky.  In the limit that the putative 51~Peg companion is dim (or
non-existent), 51~Peg itself would appear as a single star, exhibiting
visibility modulus (and trivially, visibility squared) given in a
uniform disk model by:
\begin{equation}
V^2 = (V)^2 = 
\left( \frac{2 \; J_{1}(\pi B \theta / \lambda)}
            {\pi B \theta / \lambda} \right)^2
\label{eq:V2_single}
\end{equation}
where $J_{1}$ is the first-order Bessel function, $B$ is the projected
baseline vector magnitude at the star position, $\theta$ is the
apparent angular diameter of the star, and $\lambda$ is the
center-band wavelength of the interferometric observation.  However,
if the putative 51~Peg companion were in fact luminous enough to be
detected by the interferometer, the expected squared visibility in a
narrow bandpass would be given by:
\begin{equation}
V^{2} = \frac{V_{1}^2 + V_{2}^2 \; r^2
	        + 2 \; V_{1} \; V_{2} \; r \;
	          \cos(\frac{2 \pi}{\lambda} \; {\bf {B}} \cdot {\bf {s}})}
	      {(1 + r)^2}
\label{eq:V2_double}
\end{equation}
where $V_{1}$ and $V_{2}$ are the visibility moduli for 51~Peg and the
putative companion alone as given by Eq.~\ref{eq:V2_single}, $r$ is
the apparent brightness ratio between the 51~Peg primary and
companion, ${\bf {B}}$ is the projected baseline vector at the 51~Peg
position, and ${\bf {s}}$ is the primary-companion angular separation
vector on the plane of the sky.

The key to detecting a companion to 51~Peg in PTI data is to reliably
determine the stability of the $V^2$ measured on 51~Peg.  Without a
luminous companion Eq.~\ref{eq:V2_single} predicts a stable value of
the $V^2$ observable on 51~Peg (with small variations due to baseline
projection effects with varying hour angle on the source).
Conversely, in the presence of a luminous companion
Eq.~\ref{eq:V2_double} predicts sinusoidal excursions in $V^2$ as the
system evolves and the Earth rotates; a three-magnitude fainter
companion would produce roughly 20\% peak-to-peak excursions in $V^2$.
A preliminary examination of data from 1996 suggested significant
$V^2$ variations in 51~Peg (\cite{Pan97}).  The PTI instrument
configuration for the 1997 observations reported here incorporates
compensation for spatially-varying instrument vibrations, as well as
spatial filtering to improve the visibility measurements, both of
which affected the 1996 data.  In the analysis presented here we have
placed an emphasis on choosing calibration sources and techniques that
minimize potential instrumental or environmental effects; namely we
have required calibration observations that are in close spatial (sky)
and temporal proximity to the 51~Peg observations.  Due to a limiting
K-magnitude of $\sim$ 5 (\cite{Colavita94}) this calibration strategy
forces us to use slightly resolved calibration sources, making the
absolute calibration of the $V^2$ difficult to determine.  In the
present work we have estimated the apparent diameter of the
calibration objects with respect to a model diameter for 51~Peg (Table
\ref{tab:calibrators}), and then assessed the $V^2$ stability of
51~Peg and its calibrators by inter-comparison.  Such a strategy can
say nothing about the actual apparent diameter of 51~Peg; we defer
this question to a separate publication.

\section{Observations}

The star 51~Pegasi and at least one nearby calibration object were
included in the PTI observing program on 18 nights from July 19
through November 23, 1997.  Because we have noted significant
systematic effects in measured visibilities over large sky
separations, in this analysis we have limited our attention to 51~Peg
data calibrated by two nearby calibrators, HD 215510 and HD 211006,
with similar K-band brightness (3.96) as 51~Peg (\cite{Campins85}).
The relevant parameters of the calibration objects are summarized in
Table \ref{tab:calibrators}.  These calibration objects show no
previous evidence of multiplicity or photometric variability, as well
as no evidence of multiplicity in our data (see below).  Both of these
objects are resolved by our long baseline, hence the absolute
calibration of our data depends on the calibrator diameters.
Apparent diameters for the calibration objects were estimated by
single-star fits to $V^2$ sequences calibrating the calibration
objects with respect to a single-star model 51~Peg with model diameter
of 0.72 $\pm$ 0.06 mas implied by $R_{51P}$ = 1.2 $\pm$ 0.1 $R_{\sun}$
(adopted by \cite{Marcy97}) and 65.1 $\pm$ 0.76 mas Hipparcos parallax
(\cite{Perryman96}).  The hypothesis fits themselves are discussed in
\S \ref{sec:single-star}.  This procedure is sufficient in a search
for luminous companions to 51~Peg, but leaves open the question of
51~Peg's apparent diameter.

\begin{table}[h]
\begin{center}
\begin{small}
\begin{tabular}{|c|c|c|c|c|c|c||c|}
\hline
Object    & Spectral & Star  & 51~Peg & Diam.~WRT \\
Name      & Type     & Mag   & Separation    & Model 51~Peg  \\
\hline \hline
HD 215510 & G6III    & 6.3 V & 3.1$^{\circ}$    & 0.85              \\
          &          & 3.9 K &                  & $\pm$ 0.06        \\
\hline
HD 211006 & K2III    & 5.9 V & 13$^{\circ}$     & 1.08              \\
          &          & 3.4 K &                  & $\pm$ 0.05        \\
\hline
\end{tabular}
\caption{1997 PTI 51~Peg Calibration Objects Considered in our
Analysis.  The relevant parameters for our two calibration objects are
summarized.  The apparent diameter values are determined by a fit to
our $V^2$ data calibrated with respect to a single-star model 51~Peg
using a model diameter of 0.72 $\pm$ 0.06 mas
(\cite{Marcy97,Perryman96}, see Table \ref{tab:FitSSHypothesis}).
\label{tab:calibrators}}
\end{small}
\end{center}
\end{table}

\begin{figure}
\epsscale{.7}
\plotone{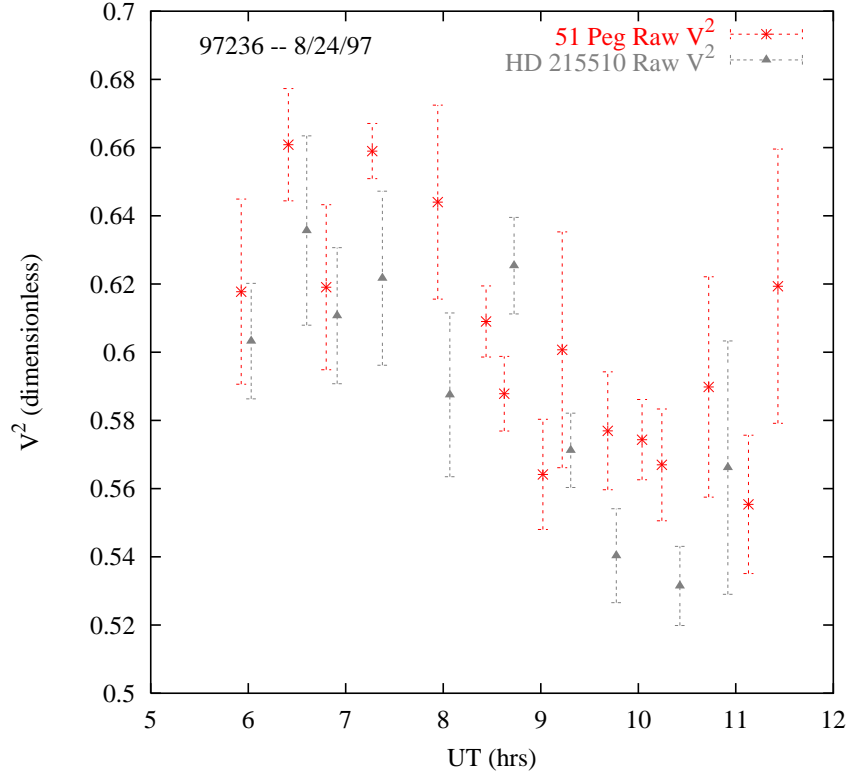}
\caption{Raw $V^2$ Data for 51~Peg and Calibrator.  This plot shows raw
$V^2$ data from a particular night (97236 -- 8/24/97) for 51~Peg and a
nearby calibrator (HD 215510, 3.1$^{\circ}$ away).  The $V^2$ data has
been averaged over the 120-second observations, and the sample
standard deviation about the mean in each observation is indicated by
the error bars.  Both 51~Peg and the calibrator exhibit formally
significant $V^2$ excursions, but both change in synchronism.  This
observation leads us to conclude that either instrumental or observing
conditions can change on time scales of approximately one hour; we
have structured our calibration procedures accordingly.
\label{fig:51praw}}
\end{figure}

Raw $V^2$ measurements were made through methods described in Colavita
(1998).  An example of the raw data from one night's (97236 --
8/24/97) observation of 51~Peg and a nearby calibrator (HD 215510) is
given in Figure \ref{fig:51praw}.

\section{Calibrated Datasets}
\label{sec:calib}
The calibration of 51~Peg $V^2$ data is performed by estimating the
interferometer system visibility using calibration sources with model
angular diameters, and then normalizing the raw 51~Peg visibility by
that system visibility estimate in order to estimate the $V^2$
measured by an ideal interferometer at that epoch.  In this letter we
consider 51~Peg datasets calibrated by the two nearby calibration
objects (Table \ref{tab:calibrators}).  We have prepared two different
calibrated 51~Peg datasets:

\begin{itemize}

\item {\it AND Dataset}: This dataset requires at least one
observation (``scan'') on both nearby calibrators within a $\pm$
one-hour calibration time window (all calibration measurements within
the time window are averaged together).  This dataset contains 105
calibrated scans on 51~Peg over 13 nights spanning a total time
interval of 123 days.

\item {\it OR Dataset}: This dataset requires at least one scan on
either of the nearby calibrators within the $\pm$ one-hour calibration
time window as above.  This dataset contains 146 scans over 18 nights
spanning the same time interval of 123 days.  As defined the AND
dataset is a proper subset of the OR dataset.

\end{itemize}

\paragraph{Calibrator Stability}

Further, as we rely on the $V^2$ stability of the two nearby
calibration objects as references for the 51~Peg analysis, it is
important to assess the relative stability of the two calibrators.
Consequently, we prepared two additional datasets for each calibration
object: calibrated with respect to the other calibration object
(i.e.~HD 215510 calibrated with respect to HD 211006 and vice versa),
and one calibrated with respect to a single-star model 51~Peg itself
using a model diameter of 0.72 $\pm$ 0.06 mas.

\section{Analysis of Calibrated Datasets}

We have analyzed the calibrated visibility datasets on 51~Peg and the
calibrators themselves by fitting single-star (Eq.~\ref{eq:V2_single})
and double-star (Eq.~\ref{eq:V2_double}) hypotheses to the datasets,
and by evaluating these hypotheses by considering goodness-of-fit
($\chi^2$) metrics.

\paragraph{Single-Star Hypothesis}
\label{sec:single-star}

\begin{figure}
\epsscale{.7}
\plotone{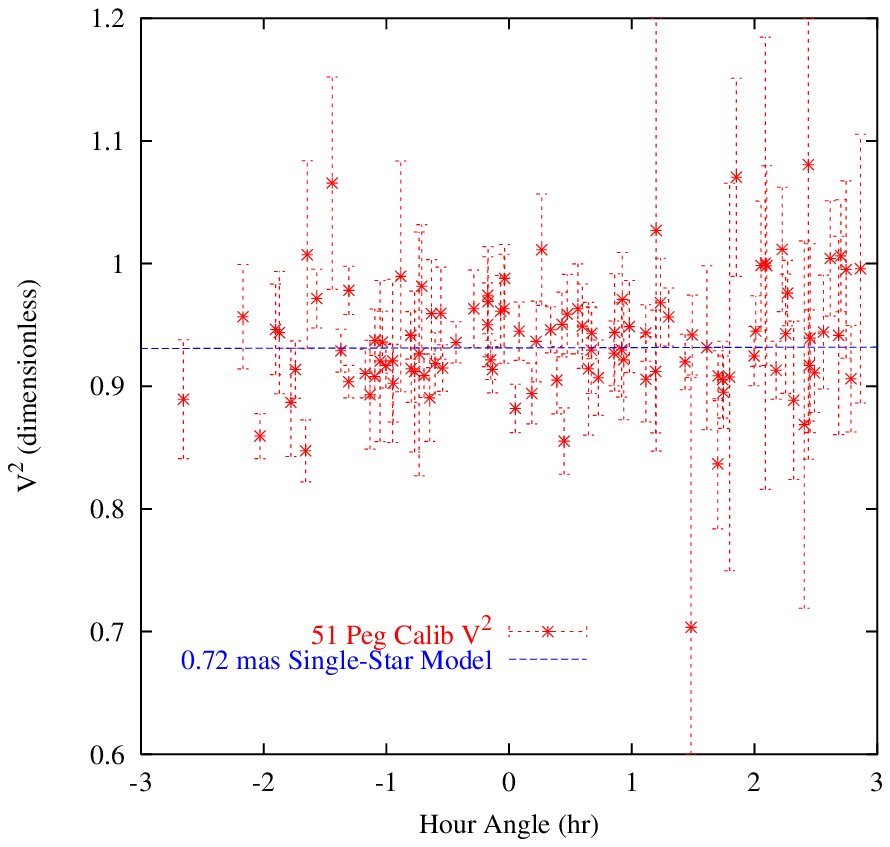}
\caption{Calibrated $V^2$ Data vs.~Hour Angle.  Assuming 51~Peg is a
single star, the $V^2$ data should follow a simple model vs.~hour
angle on the source.  This figure shows the calibrated $V^2$ from the
AND dataset, and the predicted $V^2$ vs. hour angle for a 0.72 mas
diameter single star, our model for the apparent diameter of 51~Peg
(\cite{Marcy97,Perryman96}).  The data is in good agreement with the
single-star model (see Table \ref{tab:FitSSHypothesis}).
\label{fig:51singleHA}}
\end{figure}

Since a planetary-mass companion to 51~Peg would be too dim to observe
with PTI, it is appropriate to fit a single-star hypothesis to the
calibrated datasets for 51~Peg.  To accomplish this task we have used
a global non-linear least-squares fitting code that fits a single-star
hypothesis as given in Eq.~\ref{eq:V2_single} to the input calibrated
$V^2$ datasets on 51~Peg.  The single-star hypothesis fits to our
datasets are summarized in Table~\ref{tab:FitSSHypothesis}.  The
output of the fit to the AND dataset is depicted in Figure
\ref{fig:51singleHA}, which shows a plot of the AND dataset vs.~hour
angle on 51~Peg.  For a single star the $V^2$ should follow a simple
model (Eq. \ref{eq:V2_single}).  The data exhibits good agreement with
the single-star model.

There are several notable aspects to these hypothesis fits.  The first
is to reiterate that the best fit angular diameter estimate of 0.73
$\pm$ 0.02 mas does not constitute an independent determination of the
51~Peg angular diameter -- it is just a ramification of the 0.72 mas
model diameter assumed for 51~Peg in the determination of the
calibrator angular diameters.  We further have quoted only statistical
errors on the fit diameters as determined from the internal scatter in
the $V^2$ measurements -- systematic contributions from uncertainty in
the calibrator diameters are deliberately neglected to simplify
interpretation of the $\chi^2$ results.

The second notable aspect of the single-star fits is $\chi^2$ per
degree of freedom (DOF) values that are in excellent agreement with
the expected value of 1.0.  The fit of the single-star
model is good compared to our assumed error bars based on internal
scatter of the raw $V^2$ data.  This is somewhat suprising;
while the relative weighting is reasonably well established by internal
scatter, we have no reliable a priori model for the absolute scale of
errors in our calibrated data.  The mean absolute $V^2$ residual around
the single-star hypothesis is slightly less than 3\%.  This average
absolute deviation is consistent with PTI
instrument performance in other analyses (\cite{Boden98}), the
absolute deviations seen in the calibrator data (see below), and a
good indication of the level of error in our calibrated $V^2$
measurements in a single-star model for 51~Peg.

\begin{table}
\begin{center}
\begin{small}
\begin{tabular}{|c|c|c|c|c|c|}
\hline
		& $\chi^2$ & Mean Absolute  & Fit Angular     & \#    & Calibrators \\
		& Per DOF  & $V^2$ Residual & Diameter (mas)  & Scans &             \\
\hline \hline
51~Peg AND	& 1.08	   & 0.030	    & 0.73 $\pm$ 0.02 & 105   & HD 215510, HD 211006 \\
51~Peg OR	& 1.03	   & 0.028	    & 0.73 $\pm$ 0.02 & 146   & HD 215510, HD 211006 \\
\hline 
HD 215510 Ref 1	& 1.00	   & 0.030	    & 0.86 $\pm$ 0.02 & 80    & HD 211006 @ 1.08 mas \\
HD 215510 Ref 2	& 0.65	   & 0.027	    & 0.85 $\pm$ 0.02 & 108   & 51~Peg @ 0.72 mas \\
HD 211006 Ref 1	& 0.73	   & 0.028	    & 1.08 $\pm$ 0.02 & 70    & HD 215510 @ 0.85 mas \\
HD 211006 Ref 2	& 0.54	   & 0.027	    & 1.08 $\pm$ 0.02 & 72    & 51~Peg @ 0.72 mas \\
\hline
\end{tabular}
\end{small}
\caption{Summary of Single-Star Hypothesis Fitting.  This table lists
our results on fitting single-star hypotheses to the 51~Peg and
calibrator datasets discussed in the text.  We see no evidence to
suggest an inconsistency of our data with a single-star hypothesis;
51~Peg appears as constant as our two calibration sources.  In
particular, the resulting fit diameter for 51~Peg essentially
reproduces the adopted value used to set the apparent calibrator
diameters.
\label{tab:FitSSHypothesis}}
\end{center}
\end{table}

Also contained in Table \ref{tab:FitSSHypothesis} are the results from
the calibrator inter-comparisons, and fits to calibrator datasets
using a single-star model 51~Peg as a reference.  In all cases the
agreement with single-star models is good both in an absolute
deviation ($|\epsilon|$) and a statistical ($\chi^2$) sense.  In
particular, the results in the datasets where one calibrator is
calibrating the other are consistent with the values obtained in the
51~Peg datasets.  The datasets with 51~Peg as a calibration object
actually result in fits to the calibrators that are slightly better
than the reciprocal fits to 51~Peg.  This result is reasonable, as
there are more 51~Peg scans than calibrator scans, hence the system
calibration is on average better determined using 51~Peg as a
calibrator.

In summary, our data on 51~Peg is completely consistent with a
single-star hypothesis on the scale of the observed scatter.  Further,
inter-comparison of the two calibrators yields fits to single-star
hypotheses at roughly the same level of agreement.  Nothing in our
data suggests that 51~Peg is any more variable that either of the
calibrators, both in absolute and statistical terms.

\paragraph{Binary Hypothesis}
To test the possibility of a luminous object (presumably an M-dwarf
star) as the inferred 4.2 day period companion of 51~Peg, we conducted
an experiment where we fit a binary orbit to the $V^2$ datasets,
constraining the orbit to be of the appropriate (4.231 day) period,
eccentricity (0), and approximately face-on orientation (inclination =
0 or $\pi$) to be consistent with the high-quality radial velocity
data for the system (e.g.~\cite{Marcy97}).  We performed this fitting
procedure over an input grid of semi-major axes and K-band intensity
ratios that included the values of a hypothetical M-dwarf companion in
a 4.2 day orbit.  For a given semi-major axis and intensity ratio, we
allowed the fit to solve for the optimal orbital phase parameter and
primary angular diameter.  Initial values for the angular diameters
for the primary and hypothetical secondary were set at our best-fit
single-star value, and main-sequence model value for an M3V spectral
type at the Hipparcos parallax distance, respectively.  We used this
procedure to map the $\chi^2$ surface in the subspace of semi-major
axis and intensity ratio.

\begin{figure}
\epsscale{.7}
\plotone{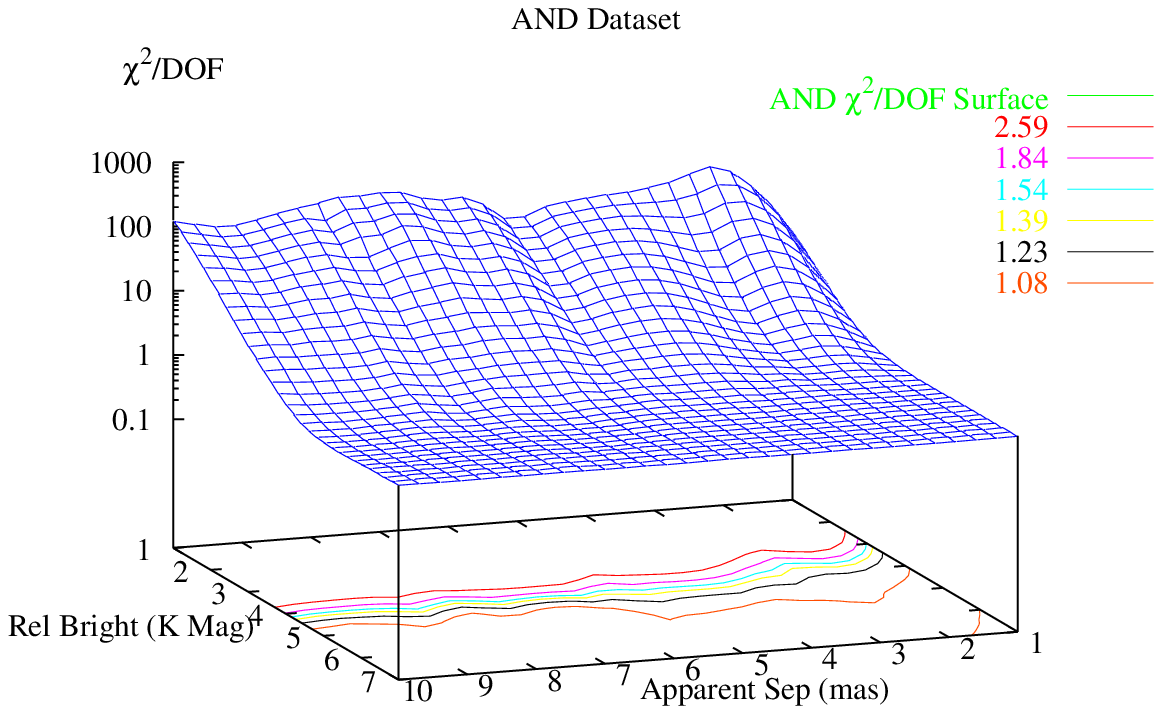}\\
\plotone{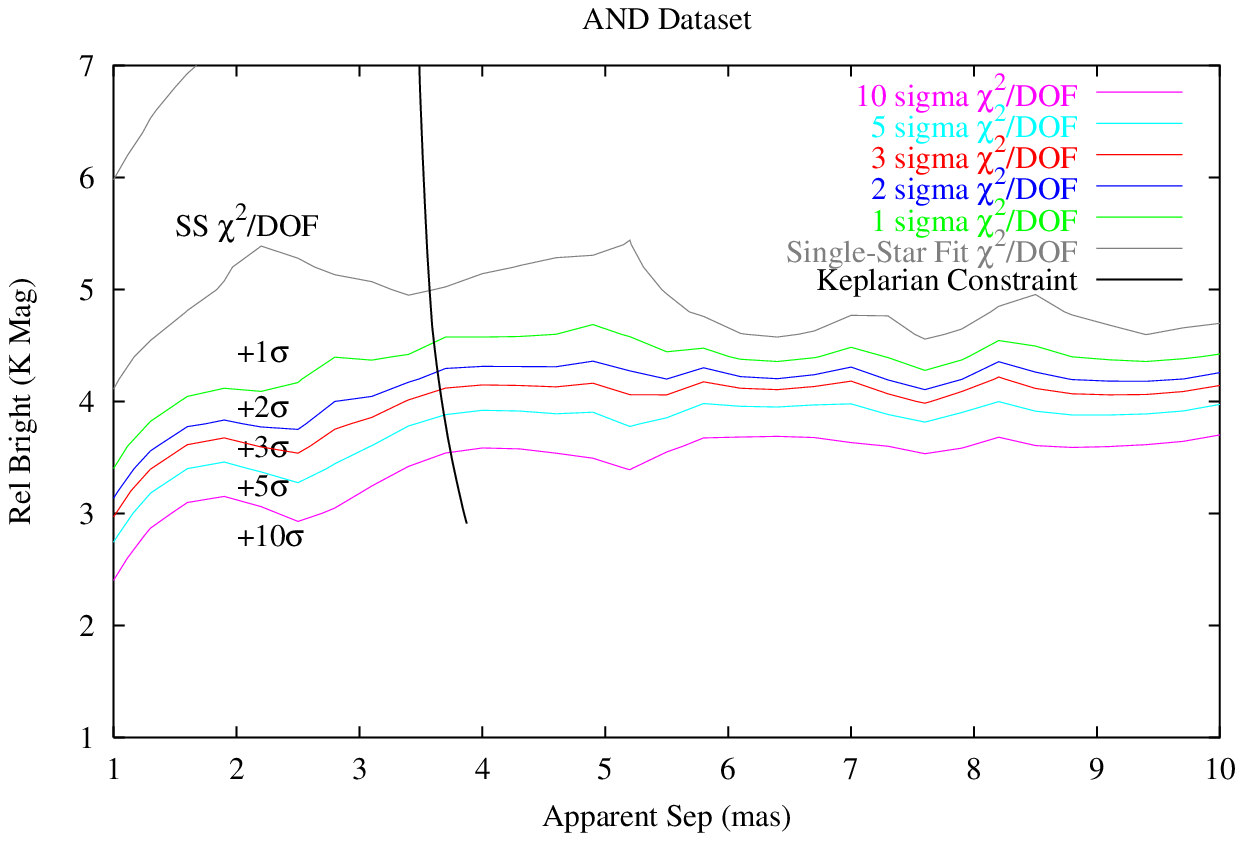}
\caption{AND Dataset Fit to a Binary-Star Hypothesis.  a (Top Pannel):
the surface of fit $\chi^2$/DOF for our AND dataset to a binary star
model for 51~Peg in the space of companion separation (in mas) and
relative K-magnitude, along with contours for the single-star
hypothesis, and +1, 2, 3, 5, and 10 standard deviations in
$\chi^2$/DOF significance.  There is no minima in this space of
companion parameters that is significantly better than the single-star
hypothesis.  b (Bottom Pannel): The contour map for the $\chi^2$/DOF
surface, and a Keplerian constraint line for a main-sequence
companion.  A Keplerian companion brighter than 4.53, 4.27, and 4.10
relative K-magnitudes to the 51~Peg primary is excluded at 68\%, 95\%,
and 99\% confidence levels respectively.
\label{fig:51binaryFit}}
\end{figure}

Figure \ref{fig:51binaryFit}a shows the result of such a fitting
procedure applied to the AND dataset.  This figure depicts the
$\chi^2$/DOF surface over values of the semi-major axis between 0.01
and 0.16 AU (projected separations between 1 and 10 mas) and intensity
ratios between 1 and 7 (K) magnitudes.  Figure \ref{fig:51binaryFit}a
shows the surface and a contour map displayed on a horizontal plane
below the surface.  At 7 magnitudes difference we are effectively
testing the single-star hypothesis against the dataset, and the binary
fit reproduces the $\chi^2$/DOF seen in the corresponding single-star
hypothesis fit.  The apparent lack of a significant minima in the
surface is striking, indicative that there is no pattern in the data
which matches the combined set of orbital constraints and a 4.2-day
period.  With decreasing relative magnitude (a brighter companion) we
see rapidly increasing fit residuals, independent of hypothetical
semi-major axis.

With 105 degrees of freedom one-sigma excursions in the $\chi^2$/DOF
around 1.0 are expected to be roughly 0.14.  Because we are uncertain
as to the absolute level of error on individual $V^2$ points, we have
scaled the $\chi^2$/DOF significance contours to match the
$\chi^2$/DOF obtained in the single-star fit; this is equivalent to
scaling the data errors to obtain a $\chi^2$/DOF of 1.0 in the
single-star fit, and allows us to compare the single-star and
binary-star models on an equal statistical basis independent of the
absolute scale of our calibrated $V^2$ errors.  Figure
\ref{fig:51binaryFit}b gives a contour map of the binary hypothesis
$\chi^2$/DOF surface with contours at the single-star fit
$\chi^2$/DOF, (scaled) contours of $\chi^2$/DOF significance, and a
constraint curve indicating a Keplerian combination of separation and
relative K-magnitude assuming a 1 M$_{\sun}$ 51~Peg and a
main-sequence M-dwarf mass/luminosity relation given by Henry and
McCarthy (1993).  The Keplerian curve intersects the 1, 2, and 3-sigma
$\chi^2$/DOF contours at 4.53, 4.27, and 4.10 relative K-magnitudes.
Assuming Gaussian errors in our data and compared with a M$_K$ for
51~Peg of 3.03 (\cite{Campins85,Perryman96}), a 4.2-day period
Keplerian companion brighter than M$_K$ of 7.56, 7.30, and 7.13 is
excluded at 68\%, 95\%, and 99\% confidence levels respectively by
this dataset.  This same analysis conducted on the OR dataset yields
slightly more stringent results (Table ~\ref{tab:FitBSHypothesis}).

\begin{table}
\begin{center}
\begin{small}
\begin{tabular}{|c|c|c|c|c|}
\hline
		& Single        & \multicolumn{3}{c|}{Companion M$_K$ Limit} \\
			        \cline{3-5}
		& Star          & + 1 $\sigma$   & + 2 $\sigma$    & + 3 $\sigma$ \\
                & $\chi^2$/DOF  &   68\% CL      &   95\% CL       & 99\% CL \\
\hline \hline
51~Peg AND	& 1.08	        & 7.56	         & 7.30            & 7.13   \\
51~Peg OR	& 1.03	        & 7.81           & 7.56            & 7.30   \\
\hline

\end{tabular}
\end{small}
\caption{Summary of Binary-Star Hypothesis Fitting.  The table gives
absolute K-magnitude lower limits for the putative 4.2-day 51~Peg
companion at nominal 1, 2, and 3 sigma significance levels, and
nominal 68\%, 95\%, and 99\% confidence levels under the presumption
of Gaussian errors in our data.  (These are to be compared with an
M$_K$ of 3.03 for 51~Peg.)  The results from all the datasets are in
good agreement, and exclude the possibility of an M-dwarf star earlier
than M5V as the putative 4.2-day period companion to 51~Peg.
\label{tab:FitBSHypothesis}}
\end{center}
\end{table}

\section{Summary}

We find no evidence to suggest that the putative 4.2-day period
companion to 51~Peg is detectable in our data; all of the datasets we
have analyzed indicate that 51~Peg is at least as stable as our two
calibration sources.  The 1997 PTI data on 51~Peg is sufficiently
stable that we can place significant limits on $\Delta$K and
consequently M$_K$ of a 4.2-day period companion.  We find upper
limits in $\Delta$K of 4.78, 4.53, and 4.27 for the 4.2-day period
companion to 51~Peg at 68\%, 95\%, and 99\% confidence levels
respectively.  These $\Delta$K limits imply companion M$_K$ limits of
7.81, 7.56, and 7.30, corresponding to upper limits on the mass of a
putative main sequence companion at 0.17, 0.20, and 0.22 M$_{\sun}$ at
the 68\%, 95\%, and 99\% confidence levels respectively
(\cite{Henry93}).  Our results cannot exclude the possibility of a
very low-mass star in a face-on orbit as the 51~Peg companion, but
such a star would have to be of spectral type M5V or later.

\end{document}